\documentclass[%
 reprint,
 amsmath,amssymb,
 aps,
]{revtex4-2}

\usepackage{aas_macros}
\usepackage{graphicx}
\usepackage{dcolumn}
\usepackage{bm}

\begin{document}

\preprint{XXX}

\title{Essay: Mapping luminous and dark matter in the Universe}

\author{Nora Elisa Chisari}

\affiliation{%
Institute for Theoretical Physics, Utrecht University, \\
Princetonplein 5, 3584 CC, Utrecht,
The Netherlands.
}%
\altaffiliation[Also at ]{Leiden Observatory, Leiden University, 
Niels Bohrweg 2, NL-2333 CA Leiden, The Netherlands.}


\date{\today}

\begin{abstract}
Our standard model of the Universe predicts the distribution of dark matter to $1\%$ at the scales needed for upcoming experiments, yet our predictions for how the luminous matter -which has interactions besides gravity- is distributed remain highly uncertain. Understanding how much gas and stars there are in the Universe and where they preferentially live is challenging, and the uncertainty affects how well we can understand the cosmological model itself. For example, it compromises our ability to tell apart different models for dark energy, the mysterious force driving the accelerated expansion of the Universe. In this Essay, I will touch upon many recent developments that suggest we will be able to overcome this limitation before data from new experiments become available. More excitingly, I will describe how our efforts to model luminous and dark matter jointly will create new possibilities for constraining the physics of supermassive black holes, galaxies, and gas over time.

\textit{Part of a series of Essays in Physical Review Letters which concisely present author visions for the future of their field.}

\end{abstract}

\maketitle

\textit{Constraining the cosmic distribution of baryons}---

We know well how {\it all} matter is distributed across much of our observable Universe. Mass maps have been produced in the last decade by several large-scale (Stage III, \cite{detf}) campaigns to survey the sky. These multi-year enterprises deliver measurements of the shapes of millions of galaxies to determine how much weak (per cent level) gravitational lensing they have suffered from all intervening matter along the line of sight. This phenomenon, predicted by the theory of general relativity \citep{Bartelmann10}, is the only direct way of mapping all matter in the Universe and how it clumps as a function of time. Almost 80$\%$ of the matter in the Universe is dark \citep{Planck}, whose nature is poorly understood \citep{Bertone}. What is left is comprised of \textit{baryons} -matter that interacts with light, like stars and gas in and outside galaxies. On large scales, we know that baryons follow the same distribution as dark matter. Hence, averaged over such scales, mass maps give us a glimpse of where {\it all} matter lies, whether dark or luminous. 

The Universe appears to be statistically isotropic and homogeneous on large scales, but it has variations in matter density on small scales. On such small scales, the assumption that baryons and dark matter clump similarly has not been a bad one so far. Because we work in Fourier space, we describe the characteristic scale of density perturbations in terms of an inverse length parameter, $k$, usually expressed in units of $h$~Mpc$^{-1}$. One Mpc $h^{-1}$ is approximately equivalent to $4.6\times 10^6$ light years. Variations of density are small and perturbative at large scales. Still, they can become large and nonlinear at smaller scales, leading to the formation of galaxies and a plethora of cosmic structures.
Predictions from numerical simulations suggest that the clumping of matter, quantified by the {\it matter power spectrum}, $P(k)$, a function of $k$, can change by typically $10\%$ and even up to $30\%$ on small scales when baryons and their corresponding physical interactions are considered.
For a long time, this difference remained undetectable. However, claims of tentative detection of a mismatch in the distribution of dark matter and baryons are beginning to emerge thanks to growing experimental precision \cite{Heymans21,Amon22,Preston23}. 
Will such a mismatch be able to alleviate the so-called {\it $S_8$ tension}~\cite{diValentino21}? The $S_8$ cosmological parameter, usually derived from measurements such as weak gravitational lensing, quantifies how much matter has clustered up to today. Recent works have shown that it deviates from the value predicted by the standard Lambda Cold Dark Matter ($\Lambda$CDM) cosmological model preferred by cosmic microwave background data.

The coming years will see the number of galaxies mapped increase from millions to billions with the start of the next generation of large astronomical surveys, i.e. the \textit{Stage IV surveys}, such as those carried out by the {\it Euclid} satellite \cite{Laureijs}, the {\it Vera C. Rubin Observatory} \cite{Rubin} or the {\it Nancy Roman Space Telescope} \cite{WFIRST}. Their goal is to map the clustering of matter with better statistics and across a longer timeline, helping address the questions of what dark matter is \cite{Bertone} and why the Universe is accelerating its expansion \cite{Weinberg,Di_Valentino_2021}. Excitingly, we can also expect them to be sensitive to differences in the clustering of baryonic and dark matter, and therefore to constrain the astrophysical processes that are responsible for it. In contrast, for these experiments, a theoretical error of $10-15\%$ in how matter is distributed could significantly bias their conclusions \cite{Semboloni11,Eifler15,Amon22,Preston23}. Measures are being taken to predict the clumping of matter at the required accuracy level. Will they be sufficient? 

In this Essay, I will first give a short review of what we do and do not know about the clustering of baryons, and describe the consequences for cosmological studies and how to mitigate them. I will then present different opportunities for simultaneously constraining the physical processes that affect the distribution of baryons across the Universe, at the same time as we explore its cosmic expansion history.    

\begin{figure}
\includegraphics[width=0.47\textwidth]{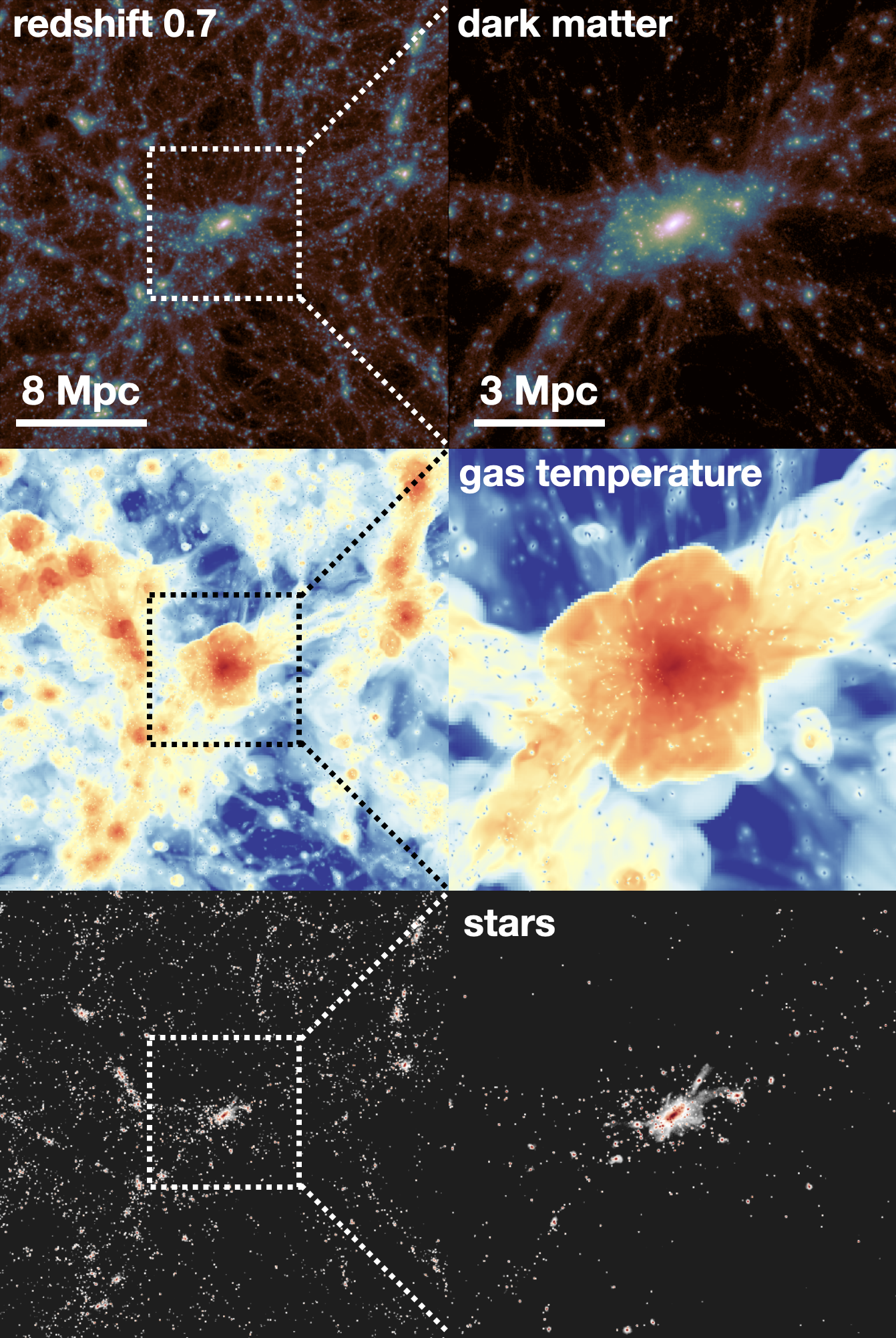}
\caption{Maps of the simulated distribution of dark matter (top), gas color-coded by temperature (middle), and stars (bottom) in the Horizon-AGN cosmological hydrodynamical simulation at redshift $z=0.7$. This roughly corresponds to the peak of lensing sensitivity for Stage IV surveys. Courtesy of Y. Dubois \cite{Dubois14}.}
\label{fig:maps}
\end{figure}

\textit{Where do baryons live?}---
It might seem incredible that we can predict the distribution of dark matter really well, while the same is not true for baryons. In the $\Lambda$CDM model, dark matter interacts only gravitationally. Modeling how it moves and clusters over time only requires capturing its gravitational interactions correctly. At sufficiently large scales, we have had the tools to predict this within general relativity for many years \citep{MaB95}. On these scales, the clumping effect of gravity can be approximated as a linear perturbation on the density of matter: $\delta(t)=\delta(t_0)D(t)/D(t_0)$, where $\delta\equiv\rho/\bar\rho-1$ is the matter density contrast, $\rho$ is the matter density at a given point and time, $\bar\rho$ is the mean matter density of the Universe at a given time, $t_0$ is today, and $D(t)$ is the growth function. On small scales, this approximation is no longer valid due to nonlinear effects, and numerical simulations are needed to determine the scale-dependent growth. $N$-body codes apply Newtonian physics -a sufficient approximation in small volumes \cite{CZ11,Hahn16}- to track the motion of dark matter particles across space and time in fictitious universes. A careful comparison of their implementation schemes has shown that they can readily predict the distribution of dark matter to within $1\%$ \cite{Schneider16}, which is comparable to the precision that will be achieved by \textit{Stage IV surveys} \cite{Huterer05} in their quest to understand cosmic acceleration.    
In contrast, although baryons represent only about $5\%$ of the total energy density in the Universe \citep{Planck}, they undergo other interactions besides gravity. Gas that was present in the early Universe cools and forms stars. Stars eject material at different stages of their lives not only through winds but also if they go supernova. Supermassive black holes, ubiquitous at the centers of galaxies \cite{Kormendy95}, trigger jets and outflows, and heat the gas around them, in what is known as Active Galactic Nuclei (AGN) feedback (Figure \ref{fig:maps}). Capturing all these processes to a reasonable degree of accuracy is required to build an unbiased picture of how matter clusters over time at small scales. Neglecting them, especially the AGN feedback, can have dire consequences on the inferred properties of the Universe particularly from gravitational lensing, e.g., \cite{Eifler15}. The goal is then clear: to enable accurate cosmological information extraction, we need to know where baryons live. 

\textit{Numerical Simulations to explore the baryonic cycle for cosmology}---
The problem of how baryons are distributed in the Universe has been recognized since the early 2000s \cite{White04,Jing06,Levine06,Guillet10}. Because the clumping of baryons is highly nonlinear and modeling requires a realistic treatment of gas physics and star formation, it remained intractable until numerous and big enough cosmological hydrodynamical simulation suites became available. In 2011, van Daalen et al. \cite{vanDaalen11} presented predictions for the matter distribution from a large suite of hydrodynamical cosmological simulations where the AGN feedback model was varied. This work demonstrated that the AGN feedback could impact the matter power spectrum by $10\%$ at $k\simeq 1\,h\,{\rm Mpc}^{-1}$ and by $30\%$ at $k\simeq 10\,h\,{\rm Mpc}^{-1}$. The net effect is a suppression of power, as opposed to the enhancement expected from gas cooling at even smaller scales, which can bias cosmological parameters up to several $\sigma$ \cite{Eifler15}. AGN feedback pushes gas out or simply prevents it from falling into dark matter overdensities by heating it. Dark matter also follows this gas due to gravitational attraction and back reacts in response to AGN feedback.

We do not know all the intricacies of the physics of star formation and feedback, nor do we have infinite resolution in time or space to deliver the perfect simulation \footnote{Simulations in small volumes tend also to be affected by lack of convergence or missing large scale power  \cite{Schaller24}.}. All simulations are an approximation, and each of them presents a different degree of {\it calibration} to existing observables (see  \cite{Kugel23} for a state-of-the-art example) - i.e., they must reproduce some quantities by design. Such quantities have most often been associated to the properties of galaxies or the stellar content of the Universe, but more recent suites use information about the gas distribution. $X$-ray observations of particularly overdense regions in the Universe known as galaxy clusters allow us to estimate the fraction of baryons in these environments. A successful simulation of the Universe should then match this observable too. 
 
\begin{figure}
\includegraphics[width=0.47\textwidth]{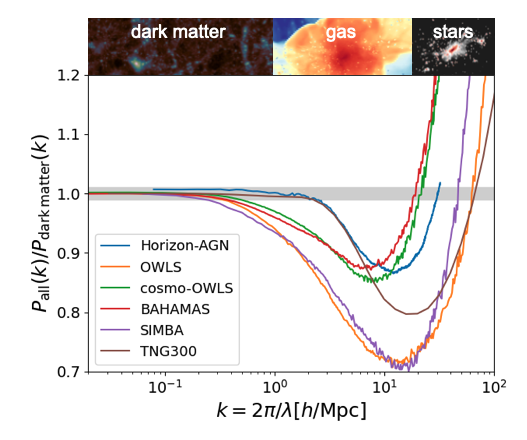}
\caption{The ratio of the matter power spectrum, which describes the clustering of matter, in hydrodynamical simulations versus dark-matter-only. On large scales, baryons follow dark matter. At intermediate scales, AGN feedback dilutes structure formation because of gas being pushed out. At small scales, the enhancement is driven by gas cooling and stars forming. The gray area represents the expected level of precision of Stage IV experiments. The curves are from the library presented in \cite{vanDaalen20}.}
\label{fig:ratio}
\end{figure}

Building on the results by van Daalen et al. \cite{vanDaalen11}, predictions for the impact of baryons on the matter power spectrum from other simulations have been steadily appearing in the literature \cite{Vogelsberger14,Hellwing16,Springel18,Chisari18,Huang19,vanDaalen20,Schaye23,Delgado23,Pakmor23}. Although these predictions share the same qualitative form, they differ in the amount of suppression produced by the AGN feedback (Figure \ref{fig:ratio}), its scale dependence, and the time evolution. These differences are larger than the expected precision of upcoming experiments, compromising our ability to extract cosmological information from smaller scales \cite{Chisari19}. 

Because of degeneracies between the parameters and resolution of numerical models, and the limited observables available for calibration, obtaining an accurate simulation of the Universe is not easy. In addition, the computational cost is significant, with typical runs requiring millions of CPU hours.
For this reason, systematically exploring the parameter space of simulated models, not to mention cosmological parameters and resolution, is unfeasible. This hinders a methodical comparison between the different simulation frameworks, which have mostly been conducted at the level of individual galaxies \cite{AGORA}. It is still an open question at what level different numerical schemes agree, with or without calibration to external observables and at different resolutions, and whether simulations are complete in the physics they are modeling.
Some teams have been able to find compromises in size or resolution, that allow them to deliver whole suites of simulations \cite{Bahamas,Quijote_sims,Camels,Salcido23,Schaye23}, occasionally aided by machine learning. One of their most interesting findings is the realization that their predictions can be unified. For example, it was found that the amount of suppression at an inverse scale of $k\sim 1\,h\,{\rm Mpc}^{-1}$ in the simulations is strongly correlated with the fraction of baryons in overdensities (halos) with masses of $M_{500,c}=10^{14}\,{\rm M}_\odot$ \cite{vanDaalen20}. The relation boasts almost no scatter and it is easy to find an empirical fit with a single free parameter. However, this desirable feature is progressively lost on smaller scales. What is then the most effective strategy to parametrize the baryonic feedback on such scales?

\textit{Modeling the clustering of baryons}---
In this context, hydrodynamical simulations are being used to test the accuracy of possible models for the distribution of baryons. Several options have become available recycling tools that were well-established for the case of dark matter. As argued before, on large scales, the gas distribution can be modeled linearly. At intermediate scales, perturbative approaches, like the effective field theory of large-scale structure (EFTofLSS, \cite{Baumann12}) are more adequate. Small scales require yet other tools, such as the {\it halo model} \cite{Seljak00,Cooray02}. The halo model assumes all matter is locked in spherical halos, whose clustering allows us to predict the large-scale, correlated distribution of gas, stars, and dark matter. Another option in this regime is a hybrid method dubbed {\it baryonification} \cite{Schneider15,Schneider19}. Baryonification is a powerful post-processing of an $N$-body simulation to account for baryons and allows one to vary the cosmology and any prior information on feedback at a lower computational cost than hydrodynamical simulations. Here, the particles in a dark-matter-only simulation are displaced to mimic the impact of feedback. The displacement is constructed with free parameters chosen to match available observations and hydrodynamical simulations.  Implemented in a small number of simulations, it can be emulated across cosmological and baryonic physics parameter space \cite{Arico21b}. Some adaptations work even at map-level \cite{Anbajagane24}.

These methodologies have advantages and limitations and must be applied differently. The EFTofLSS relies on assuming our Universe satisfies certain symmetries and General Relativity to provide a theoretical prediction that has been validated up to quasi-linear scales, including the modeling of AGN feedback \cite{Lewandowski15}. Because the EFTofLSS is not designed to be accurate in the nonlinear regime, one must be willing to lose the corresponding scales in the data (optionally, by introducing a theoretical error \cite{Maraio24}), or restrict to epochs of the Universe where the growth of structure is closer to linear \cite{Braganca21}. Interestingly, Lewandowski et al. \cite{Lewandowski15} predict that only one parameter is needed to describe baryonic feedback at intermediate scales. This might explain why other authors \cite{Mohammed14,Joudaki18,vanDaalen20,Kokron21} have found a one-parameter model to also be effective in describing the output of simulations at $k\lesssim 0.6-1\,h\,{\rm Mpc}^{-1}$. Remarkably, this means that the free parameters of different models should be correlated.

Can we really encapsulate the distribution of baryons in the Universe with a single parameter? Feedback is a complicated process with perhaps more than one mechanism at play \cite{Oei24} and a possible dependence on cosmology \cite{Pranjal24}. It has admitted one-parameter models so far probably because the data is only on the verge of constraining the effect. Depending on the nature of the suppression, we might already be seeing a bias in cosmological constraints from gravitational lensing \cite{Bigwood}. The next generation of experiments will detect the effect of feedback at high significance (e.g., Ref.\cite{6x2pt}) and this will likely drive the need for more flexibility in the modeling. The redshift evolution of the feedback strength might also vary with time nontrivially, requiring additional freedom. 

The halo model \cite{Fedeli,Fedeli2,Semboloni11,Mead,Mead2,Mohammed14} indeed requires more parameters to describe the connection between baryons and dark matter at small scales, even at fixed time. In this model, matter is composed of dark matter, stars, and gas distributed in spherically symmetric over-densities across the Universe. Some schemes also include a diffuse gas component \cite{Fedeli,Schneider19}. For application to gravitational lensing, resolving the distribution of stars is unnecessary. One can consider them to be infinitely concentrated at the center of halos. The fraction of stars in each halo -the {\it stellar-to-halo mass relation} \cite{Behroozi10,Moster13}- needs to be known to complement the model. This can be obtained self-consistently from gravitational lensing \cite{Zu15,Dvornik23}. 
Within the halo model, the challenge lies in bridging small and large scales \cite{Mead21}. What would be an accurate model of this transition scale, and is it connected to our inherent definition of a halo \cite{Debackere,Salazar24}? These and many other modeling challenges start to emerge as more and more precise observational probes are brought into the picture.

\textit{Observational probes to explore cosmic baryons}---
Modeling is advancing at full speed. Simulations, despite their costs, are also becoming more widely available. What are we overlooking in addressing the issue of baryons in the coming decade? 

Recent studies \cite{Debackere,vanLoon24} have argued that we are limited by our current knowledge of how baryonic processes affect halos hosting groups of galaxies. For example, we have a limited understanding of how often and for how long AGN are active, i. e. their \textit{duty cycle} \cite{Eckert21}. They further suggested that at scales lower than $k\sim 3\,h\,{\rm Mpc^{-1}}$, even lower mass halos become dominant \cite{vanLoon24}.
All these pieces of evidence indicate that the ideal partner to cosmological weak lensing are observations of the baryonic content of intermediate and low-mass halos. In addition, these observations should probe the epoch of the Universe where lensing has the most sensitivity for upcoming surveys (around $t\sim 8\,{\rm Gyr}$). 
Obtaining such observations is nontrivial. Fortunately, there are several direct and indirect options available, along with new emerging methods that I will now describe.

Traditionally, gas in groups has been elusive because of its low surface brightness in the $X$-rays. Attempts to constrain the gas content of groups of galaxies directly have been limited to the present epoch and are sensitive to selection cuts \cite{Sun09}. Therefore, it is still unclear whether these samples, widely used in the calibration of baryonic effects, are representative. Mock observations have suggested that selection biases will persist because of inherent scatter in the $X$-ray brightness of groups \cite{Marini24}. In other words: It is possible that we might be overestimating the mass of gas in groups. Even if samples are selected from the abundance of galaxies, modern $X$-ray observatories are still limited to a smaller cosmological volume compared to weak lensing \cite{Kluge24}. In consequence, significant efforts are ongoing to understand $X$-ray emission in galaxy groups and their role in calibrating baryonic effects in cosmology \cite{Akino22,Grandis24,Eckert24,Popesso24}.  

The gas surrounding galaxies, groups, and clusters is hot and ionized, and it moves due to gravity. This imparts a Doppler effect on backlight photons from the cosmic microwave background - the {\it kinetic Sunyaev-Zeldovich (kSZ) effect} \cite{kSZ72,kSZ80,Ostriker86}. In this way, the kSZ effect directly probes the gas content of the Universe. Compared to $X$-ray selected samples, it has the advantage of being immune to selection effects. The first kSZ analysis showing that the way baryons cluster has a significant impact on how we interpret lensing measurements was performed by \cite{Amodeo21} using data from the {\it Atacama Cosmology Telescope}. More recent studies \cite{Bigwood,Schneider22,McCarthy24} used the baryonification model to perform a joint analysis of current weak lensing and kSZ observations and concluded that existing baryonic feedback prescriptions are not strong enough to reproduce these observations jointly. This could potentially give an astrophysical explanation for the $S_8$ tension. With the advent of {\it Simons Observatory} \cite{Ade19}, the expectation is for the kSZ effect to play a prominent role in constraining the amount of gas in low-mass halos in the near future \cite{Battaglia17,Ade19}.

To directly trace hot gas in the Universe, other studies  \cite{Troester22,Sanchez23} have relied on the {\it thermal Sunyaev-Zeldovic (tSZ)} effect, a distortion of the cosmic microwave background temperature that arises when photons scatter off the electrons in the hot gas. These studies align more closely with simulations involving strong feedback scenarios: According to these measurements, more gas needs to be pushed out of halos, and to a larger distance. Could there be alternative explanations? For example, is this conclusion affected by the specifics of the modeling of the tSZ signal or by the simulation considered \cite{Grayson23}? Alternatively, could a preference for a higher suppression in the matter power spectrum be explained by a change in the best-fit cosmological model? If so, tSZ measurements would confirm the $S_8$ tension \cite{McCarthy23}. 

The aforementioned tSZ measurements are an example of a \textit{cross-correlation} measurement. Maps of the Universe sensitive to similar physics share some common traits. These can be extracted through their cross-correlation - an operation where patterns in the maps are cross-examined for commonalities. For two three-dimensional scalar fields $F_1(\bar{x})$ and $F_2(\bar{x})$, the correlation function is given by: $\xi_{12}({\bf r})=\langle F_1(\bar{x}) F_2(\bar{x}+{\bf r})\rangle$. A trivial example is the correlation of the density field with itself. The Fourier transform of this correlation is the matter power spectrum. In addition to scaling relations, cross-correlations of weak lensing maps with other observables can constrain the amplitude of baryonic effects and the distribution of the gas without selecting specific objects. These objects form part of the map, but there is also valuable information about objects that have not been selected. 

Another example of a cross-correlation measurement to further constrain the gas distribution in the Universe was recently presented in \cite{Ferreira24}. Here, the cross-correlation was performed between $X$-ray maps from the {\it ROSAT} satellite and weak lensing from the Dark Energy Survey. The measurement has a very high signal-to-noise. Tentatively, this approach yields a higher power spectrum suppression compared to studies that rely on individual cluster or group measurements.

Cross-correlations with other weak lensing surveys and new $X$-ray data \cite{Lau24} should be possible in the near future. The increased signal-to-noise could drive a major improvement in our knowledge of AGN feedback, and an additional point of comparison for numerical simulation. At the same time, interpreting these measurements will likely become more challenging. A recent work that combines $X$-ray and tSZ cross-correlations with weak lensing \cite{LaPosta24} showed that the $X$-ray emission from AGN themselves must be accounted for, and that including nonthermal contributions to the tSZ signal also brings the model into better agreement with the data. This begs the question of whether our current models for these observables are accurate enough in light of the upcoming generation of experiments.

Another promising direct avenue to constrain the distribution of baryons is through {\it fast radio bursts} (FRBs) \cite{Lorimer18,Macquart18,Macquart20}, short pulses in the radio coming from extragalactic sources. There are about 800 known FRBs so far; their origin is unclear, but we know that as these pulses fly through the Universe, they traverse ionized gas. The number density of electrons in the intergalactic medium is estimated by the dispersion measure of the pulse at arrival. This information can be used directly to complement weak lensing observations \cite{Reischke23}. With new facilities, we can expect the number of these mysterious transients to grow to $\sim 10^4$ per decade. Can consistent models for $X$-rays, kSZ, tSZ and FRB observables be developed? Such models would be key to paint a complete picture of the distribution of gas in the Universe.

An emerging method to indirectly probe the impact of AGN feedback on the composition of the Universe is through its impact on the cosmic star formation history \cite{Jego23,Yan24}. It heats up the gas in dense halos, which in turn suppresses star formation. As a result, measurements of the mass density in stars over time are likely sensitive to the strength of AGN feedback due to this dynamic \cite{Hopkins06,Behroozi13,Madau14}. Yan et al. \cite{Yan24} investigated the cosmic mass density in stars by analyzing the cross-correlation of the infrared light emitted by galaxies throughout cosmic history, known as \textit{cosmic infrared background} (CIB), alongside galaxy maps. Their findings align with constraints obtained from multi-wavelength surveys. In addition to examining stellar content, the variety of galaxy morphologies and their scaling relations may also contribute to our understanding of AGN feedback \cite{Dubois16,Rosito21,Byrne24}. It remains an open question how much constraining power can be harnessed from these indirect probes of AGN feedback.

\textit{Beyond two-point statistics.}--- Stage IV weak lensing surveys will be extremely powerful, enabling a model-independent reconstruction of the matter power spectrum \cite{Preston24,Broxterman24b}. However, two-point correlations provide a complete way to compress information only when the fields being correlated are Gaussian distributed. This condition holds true for the Universe's density field in its early stages. Over time, however, gravity and other nonlinear processes break this assumption. This means that the two-point statistics must be complemented by other measures. Many examples of going beyond two-point statistics exist \cite{Yang13,Semboloni13,Paillas17,Weiss19,Asgari20,Foreman20,Grandon24,Srinivasan24, Broxterman24,Saha24}. The significant potential they offer for understanding cosmological parameters and AGN feedback justifies the effort to pursue them. Ambitious plans are starting to be realized to infer the cosmological model alongside AGN feedback, even at the field level \cite{Porqueres23}. 

Finally, beyond two-point statistics of observables other than weak gravitational lensing can also competitively constrain the impact of feedback on the distribution of baryons. Sky-averaged spectral distortions of the cosmic microwave background blackbody curve probe the ionized gas content of the Universe and they have been shown to have high sensitivity to baryonic feedback scenarios \cite{Hill15,Thiele22}.

\textit{Concluding remarks.}---
There is no shortage of options for constraining the distribution of baryons across the Universe. In general, a variety of existing and new observables can be used to tackle this problem. 
Combining them in a robust manner presents a challenging but achievable goal. By employing these observables more holistically and in combination, we will test the consistency of their models, we will test the consistency of their models, and inform simulated models in cosmological simulations further. This feedback loop between observations and modeling will ultimately enable us to harness more efficiently the power of our data for cosmology and will lead to a comprehensive view of gas, black hole, and stellar formation physics across cosmic time. 

\begin{acknowledgments}
I would like to thank the Instituto de F\'isica de Altas Energ\'ias (IFAE) in Barcelona, Spain, and the Institute for Advanced Study and Princeton University, Princeton, USA, for their hospitality while this work was in progress. I also thank the organizers of the 2024 Mock Barcelona conference for useful discussions that helped inform this work. I thank Yohan Dubois for providing Figure \ref{fig:maps} based on the Horizon-AGN data \cite{Dubois14}. I also thank him, Simone Ferraro, Alex Amon, David Alonso and Matthieu Schaller for comments that helped improve this Essay. 
\end{acknowledgments}

\begin{figure}
    \centering
    \includegraphics[width = 0.7\columnwidth]{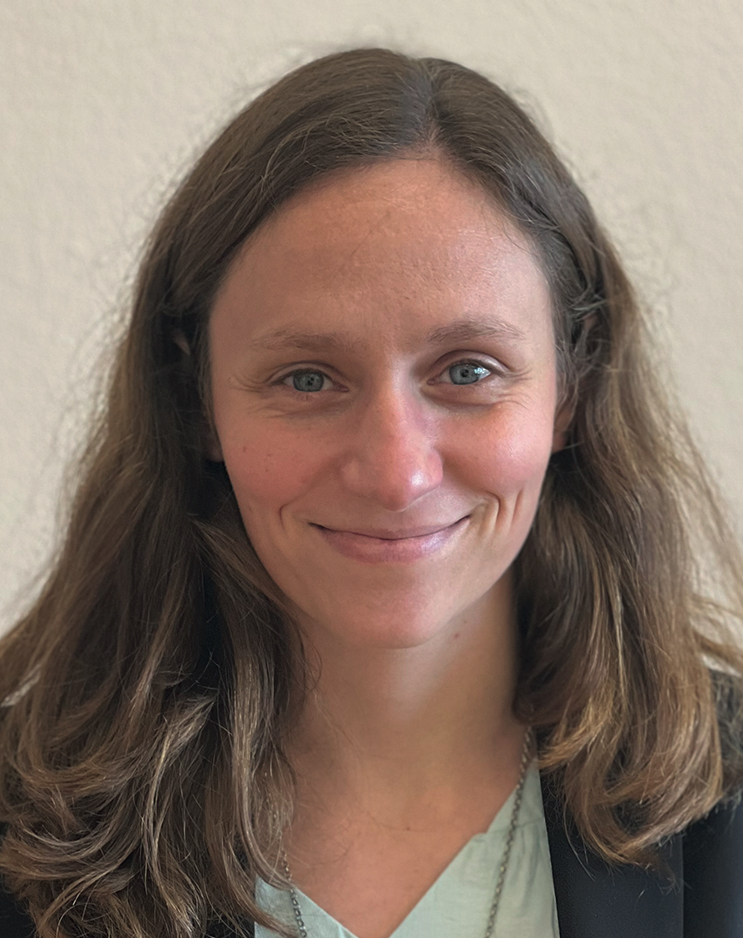}
    \caption{Elisa Chisari is a cosmologist and assistant professor at Utrecht University, Netherlands. She received her Ph. D. in astrophysics in 2014 from Princeton University and is expert in gravitational lensing, baryonic effects, and intrinsic alignments. She works at the intersection of theoretical and observational cosmology, modeling the large-scale structure for application in galaxy surveys. Dr. Chisari is member of the Kilo-Degree Survey and the current deputy analysis coordinator for static probes of the LSST DESC collaboration. She has been recipient of a Royal Astronomical Society Research fellowship (UK) and an NWO Vidi grant (The Netherlands) to support her research.}
    \label{fig:figure1}
\end{figure}

\bibliography{main}

\end{document}